\documentclass{article}

\usepackage{PRIMEarxiv}

\usepackage[utf8]{inputenc} 
\usepackage[T1]{fontenc}    
\usepackage{hyperref}       
\usepackage{url}            
\usepackage{booktabs}       
\usepackage{amsfonts}       
\usepackage{nicefrac}       
\usepackage{microtype}      
\usepackage{lipsum}
\usepackage{fancyhdr}       
\usepackage{graphicx}       
\graphicspath{{media/}}     
\usepackage{cite}
\usepackage{amsmath,amssymb,amsfonts}
\usepackage{balance}
\usepackage{algorithmic}
\usepackage{tabularx}
\usepackage{caption}
\usepackage{textcomp}
\usepackage{xcolor}
\usepackage[numbers]{natbib}
\def\BibTeX{{\rm B\kern-.05em{\sc i\kern-.025em b}\kern-.08em
   T\kern-.1667em\lower.7ex\hbox{E}\kern-.125emX}}

\title{Robotising Psychometrics: Validating Wellbeing Assessment Tools in
Child-Robot Interactions

}
\author{
  Nida Itrat Abbasi\thanks{\textit{\textbf{\scriptsize{\underline{Nida Itrat Abbasi and Guy Laban have contributed equally to this work and share first authorship, and are also the corresponding authors.}}}}} \\
  Department of Computer Science and Technology \\
  University of Cambridge \\
  Cambridge, UK\\
  \texttt{nia22@cam.ac.uk} \\
   \And
Guy Laban$^{*}$ \\
  Department of Computer Science and Technology \\
  University of Cambridge \\
  Cambridge, UK\\
  \texttt{gl538@cam.ac.uk} \\
   \And
  Tamsin Ford \\
  Department of Psychiatry \\
  University of Cambridge \\
  Cambridge, UK\\
  \texttt{tjf52@medschl.cam.ac.uk} \\
     \And
  Peter B. Jones \\
  Department of Psychiatry \\
  University of Cambridge \\
  Cambridge, UK\\
  \texttt{pbj21@cam.ac.uk} \\
     \And
Hatice Gunes \\
  Department of Computer Science and Technology \\
  University of Cambridge \\
  Cambridge, UK\\
  \texttt{hg410@cam.ac.uk} \\
}

\begin{document}
\maketitle

\begin{abstract}
\small{The interdisciplinary nature of Child-Robot Interaction (CRI) fosters incorporating measures and methodologies from many established domains. However, when employing CRI approaches to sensitive avenues of health and wellbeing, caution is critical in adapting metrics to retain their safety standards and ensure accurate utilisation. In this work, we conducted a secondary analysis to previous empirical work, investigating the reliability and construct validity of established psychological questionnaires such as the Short Moods and Feelings Questionnaire (SMFQ) and three subscales (generalised anxiety, panic and low mood) of the Revised Child Anxiety and Depression Scale (RCADS) within a CRI setting for the assessment of mental wellbeing. Through confirmatory principal component analysis, we have observed that these measures are reliable and valid in the context of CRI. Furthermore, our analysis revealed that scales communicated by a robot demonstrated a better fit than when self-reported, underscoring the efficiency and effectiveness of robot-mediated psychological assessments in these settings. Nevertheless, we have also observed variations in item contributions to the main factor, suggesting potential areas of examination and revision (e.g., relating to physiological changes, inactivity and cognitive demands) when used in CRI. Findings from this work highlight the importance of verifying the reliability and validity of standardised metrics and assessment tools when employed in CRI settings, thus, aiming to avoid any misinterpretations and misrepresentations.}
\end{abstract}

\keywords{Child--robot Interaction \and Wellbeing \and Validity \and Reliability \and Psychological Assessment \and Social Robots}

\section{Introduction}
Mental health and clinical psychology research has been exploring the use of novel technologies in diagnosis, assessment, and treatment. The use of automated `machine-learning' based approaches has been gaining traction \cite{dwyer2018machine}, especially interactive artificial agents such as social robots since they have been successfully employed to promote self-disclosure and elicit information among research participants and patients alike \cite[e.g.,][]{abbasi2022can,Laban2021,Laban_blt_2023,uchida2017robot}.
Robots allow individuals to confide about themselves with minimal social consequences \cite{sharing_2023,open2023}, thereby having the potential to alleviate the pressure on mental health services that currently suffer from extended wait times \cite{Laban2022}. Nevertheless, researchers need to be mindful of ``\textit{robotising}" psychological assessment tools and paradigms, as these instruments have been validated based on specific assumptions and for particular modes of administration \cite{baxter2016characterising}. They must ensure that measures are being utilized accurately, especially when evaluations might be performed automatically using computational tools. 

Moreover, developing robots for different domains and settings requires discipline-specific insights that need to be accounted for, to avoid potential mismatches in adaptations during the transferability of skills from one field to another \cite{vsabanovic2009outside}. Traditionally, confirming the appropriate use of a specific metric, methodology, or assumption involves a thorough assessment of the underlying construct and its operational considerations, ensuring the validity and reliability of its application in research and practice. Therefore, adaptations to `robotise' established tools from health and wellbeing research require further validation, to ensure they are both methodologically rigorous and ethically implemented. Thus, researchers can confirm that the measures meet their intended purpose \cite{baxter2016characterising} and fit the novel settings. 

This paper presents the \textit{first steps} towards investigating the transferability of established psychometric measures like the Short Moods and Feelings Questionnaire (SMFQ) and the Revised Child Anxiety and Depression Scale (RCADS) for assessing the wellbeing of children using child-robot interaction (details of the dataset mentioned in \cite{abbasi2022can}). At first, we examined the reliability of these tools to determine their susceptibility to random errors when communicated by a social robotic agent (Nao by SoftBank Robotics). Subsequently, to assess the construct validity of these tools when transferring them to CRI experimental environments, we performed a confirmatory principal component analysis (PCA) for each measure.
We aim to provide a critical perspective, showing that when applying assessment methods in interdisciplinary settings like CRI, researchers should err on the side of caution to avoid misrepresentation or mismatching of metrics. 
 
\section{Related Works}
\subsection{Wellbeing assessments for children}

Understanding children's wellbeing is integral to enhancing their mental health. As such, several initiatives have been employed, that use established psychological questionnaires typically measuring anxiety and depression, to identify wellbeing challenges and emotional concerns at an early stage \cite{ford2020data,mansfield2020oxwell}.
For example,  the Moods and Feelings Questionnaire (MFQ) has been used to assess and monitor symptoms of depression in children between the ages of 6 to 19 years by utilising about 33 items that measure children's feelings over the last two weeks\footnote{https://www.corc.uk.net/outcome-experience-measures/mood-and-feelings-questionnaire-mfq/}. A condensed version, SMFQ consisting of 13 items, can also be used to gauge children's behaviour over the same time frame \cite{angold1995development}. Similarly, RCADS is another standardised and validated questionnaire, consisting of 6 subscales that have been used to assess anxiety and depression in children from the ages of 8 to 18 years \cite{chorpita2005psychometric}.

Traditionally, these measures followed a paper and pen based mode of administration that required children to respond by circling the responses that they felt were representative of their feelings \cite{angold1995development,chorpita2005psychometric}. However, currently, surveys are predominantly employing web-based forms that require children to select their responses on a computer screen \cite{mansfield2020oxwell}. In recent times, application-based programs that employ computerised adaptive testing have been gaining popularity because of their time efficiency \cite{stochl2021modernising}. For example, Artemis-A utilises items from standardized wellbeing questionnaires while at the same time eliminating the need for children to respond to every item in order to assess their wellbeing\footnote{https://www.artemis-a.org/}. This could also pave the way for robotised adaptive testing that blends the time efficiency of programs like Artemis-A with the friendly appearance and behaviour of robots.

\subsection{Applied psychological methods for HRI and CRI}

By adapting, \textit{transfering}, and recontextualizing psychological methods, researchers across various disciplines are answering existing questions more effectively and uncovering new realms of inquiry. The integration of psychological methods is particularly meaningful in HRI, a field which is leveraging psychological principles to foster more natural and effective interactions between humans and robots \cite{Henschel2021}. Therefore, adapting psychological methods is essential to address the distinct challenges and dynamics of HRIs. 

Behavioural experiments are reconfigured to include robot-specific factors such as design, autonomy, and interactivity, thereby understanding human responses to different types of robotic behaviours and appearances. For example, a previous study \cite{anna_collabra_2021} adjusted the classic Stroop paradigm \cite{Stroop1935} to asses social perception of robots using cognitive and behavioural measures. To fit in HRI settings the task manipulation changed to demonstrate robotic faces instead of typical manipulations employed via Stroop tasks \cite[see][]{Stroop1935}. Perez et al.  \cite{perez_2018} adapted gaze cueing paradigms to assess the human's reception of a robot in HRI protocols and demonstrated a similar pattern of data to what has been previously observed in typical cognitive psychology experiments. Finally, Di Nuovo et al. \cite{di2019assessment} have also adapted the standardised paper and pen-based Montreal Cognitive Assessment for an HRI setting, using the Pepper robot and the IBM Watson cloud services, to screen for mild cognitive impairments among their participants.

Psychometric tools and measurements, crucial in psychological assessments, are customized for HRI to quantify social, behavioural, cognitive and performance-related aspects \cite[see][]{RefWorks:486}. These adaptations are not just operational but also conceptual, ensuring that these tools yield insights that are both relevant and impactful for the design, development, and evaluation of HRI. While some metrics and measurements have been developed specifically for HRI contexts (such as ROSAs \cite{Carpinella2017} \& Goodspeed \cite{RefWorks:162}), HRI research often relies on adapting established scales, adjusted to be applied in HRI settings. For instance, the Jourard Self-Disclosure Questionnaire \cite{RefWorks:509} is used to assess individuals' perception of their self-disclosure towards human counterparts in daily environments by retrospectively reflecting on such instances. This scale has been successfully modified in previous HRI studies to evaluate individuals' perception of their self-disclosure to robotic agents in experimental settings \cite{Laban2021,Laban_blt_2023}.

Amidst growing recognition of open science initiatives and the replication crisis \cite{Gunes2022,Henschel2021}, researchers are dedicating efforts to replicate human-human behavioural paradigms into the realm of HRI \cite[e.g.,][]{anna_collabra_2021,perez_2018}, and undertaking further assessments of their HRI paradigms via direct replication \cite[e.g.,][]{Laban2021}. Nonetheless, HRI studies rarely conduct a comprehensive evaluation of methodological transferability, diagnosing the reliability and validity of human psychology methods to HRI.

\section{Methodology}

\subsection{Transferability analyses}

We inspected the reliability and construct validity of SMFQ and RCADS within a CRI context. Assessing reliability in HRI and CRI contexts is critical for detecting and addressing medium-specific random errors. Initially, the overall scales' reliability was assessed using Cronbach's \(\alpha\), to indicate a general reliability score. Subsequent diagnostic steps involved standardizing items, evaluating scale's means, variances, and item-total correlations, and evaluating the scale's performance with hypothetical item deletions to understand individual items' influence on the overall scale variance and consistency. These steps collectively assessed the scale's internal consistency and item contributions to the overall measurement of the construct.

For assessing the construct validity a Confirmatory PCA was conducted for identifying the underlying factor structures and determining the feasibility of these measures when applied in CRI settings when a robot communicates scale items.  Initially, a correlation matrix is inspected to ensure no multicollinearity or threats to discriminant validity, ensuring that the items are distinct. The Kaiser-Meyer-Olkin (KMO) measure and Bartlett’s test of sphericity confirm the data's suitability for PCA, indicating substantial correlations among items. The Measures of Sampling Adequacy (MSA) and communalities assessment further justify PCA by showing each item shares sufficient variance with others and indicating the proportion of variance each item contributes to the principal component. Finally, the PCA identifies a principal component explaining a significant portion of variance, with item loadings supporting the unidimensionality of the scale. High loadings indicate an item's strong relevance to the construct, while low loadings signal the need for reflection about the tool's effectiveness in CRI. This systematic approach ensures the data's appropriateness for PCA and reveals the underlying factor structure of the scale's items. The analysis extended to a self-reported version of the RCADS, aiming to uncover potential differences introduced by the medium of data collection, thereby offering a comparative insight into how the mode of administration influences psychological assessments.

\subsection{Summary of CRI study}
\label{SetupTasks}
In a previous study, 41 participants\footnote{We collected data from 13 more participants after the 28 participants mentioned in \cite{abbasi2022can} to balance the gender across the study.} between the ages of 8-13 years old interacted for about 30-45 minutes with a Nao robot in a laboratory setting \cite{abbasi2022can,abbasi2023humanoid}. In this work, we used the data collected from 36 participants (mean age = 9.44, std dev = 1.36; 19 boys and 17 girls) as their respective parents/guardians had consented to data sharing with a new researcher who recently joined the team and is part of this work. The robot followed a pre-programmed script and conducted 5 tasks throughout the session: (1) Ice-breaking task (Task 0); (2) Recalling a recent happy memory and a sad memory (Task 1); (3) SMFQ (Task 2)\footnote{https://www.corc.uk.net/outcome-experience-measures/mood-and-feelings-questionnaire-mfq/}; (4) Picture task inspired from the Child Apperception Test (CAT) (Task 3); and (5) three subscales of RCADS (Task 4)\footnote{ https://www.corc.uk.net/outcome-experience-measures/revised-childrens-anxiety-and-depression-scale-rcads/}. Further details of the experiments can be found in \cite{abbasi2022can}. 

\subsection{Measures}
Since this work investigates the transferability of established psychometric measures for the assessment of mental wellbeing in CRI settings, we have primarily focussed on Task 2 i.e. the SMFQ \cite{angold1995development} task and Task 4 i.e. the RCADS \cite{chorpita2005psychometric} task mentioned above. Missed/incorrect responses were inputted with the mode value of the response ratings for each child for both scales.\\
\textbf{(i) SMFQ:} The SMFQ consisted of 13 items that were verbalised by the robot sequentially, to measure the mood of the child in the last two weeks. For example, item 11 included the statement ``You felt that nobody really loved you". The child could respond verbally with ``Not true", ``Sometimes" or ``True". Each response rating corresponded to a numerical score: Not true = 0, Sometimes = 1, and True = 2. The robot would continue with the next item of the questionnaire after a wait time of about 5-6 seconds. \\
\textbf{(ii) RCADS:} The RCADS tasks consisted of the robot verbalising sequentially the items of three subscales: generalised anxiety (6 items), panic (9 items) and low mood (10 items). For example, item 2 included the statement, ``You worry that something awful might happen to someone in your family". The child could respond verbally with ``Never", ``Sometimes", ``Often", or ``Always". The numerical scores for each response ratings were: Never=0, Sometimes=1, Often=2, and Always=3. After a wait time of 5-6 seconds, the robot would continue with the next item on the questionnaire. The self-report version of the questionnaires was sent out as online forms and children were requested to complete them prior to their experiment session.\\

\section{Results}
\subsection{Reliability Analysis}

\subsubsection{SMFQ}

The SMFQ administered by a robot demonstrated strong reliability, with Cronbach's \(\alpha\), score of .85 for the overall scale (also observed in our previous work \cite{abbasi2023humanoid}). When the items were standardized, the reliability increased marginally to .88, indicating that the scale items are consistently measuring the construct of interest across different individuals. The summary item statistics indicated that the mean of the item means was .35, with a minimum item mean of .08 and a maximum of .81, yielding a range of .72. This maximum to minimum ratio of 9.67 reflects a relatively high spread in the central tendency of individual item scores. Item variances had a mean of .34, ranging from .08 to .66 with a range of .58, which signifies a moderate level of variability within item responses.

Item-total statistics revealed the scale mean if an item was deleted ranged from 3.72 to 4.44, suggesting that no single item disproportionately influenced the overall scale scoring. The scale variance if an item was deleted displayed modest variation across the items, indicating that the individual items contributed evenly to the scale variance. Corrected item-total correlations ranged from .27 to .79, signifying that most items correlated well with the total score, although some items showed a lower correlation. This could imply that certain items may not align as closely with the overall construct being measured. Cronbach's \(\alpha\) if an item was deleted ranged from .83 to .87, indicating that the internal consistency of the scale remained high and relatively unaffected by the removal of any individual item. These values suggest that the scale would retain its reliability even with the potential exclusion of specific items. The scale statistics showed an overall mean score of 4.53 with a variance of 20.71 and a standard deviation of 4.55 across the 13 items. This points to a relatively high dispersion of overall scores among the sample.

\subsubsection{RCADS}
\paragraph{Robot-Administered}
The RCADS administered by a robot demonstrated strong reliability, with Cronbach's \(\alpha\), score of .92 for the overall scale. When items were standardized, the Cronbach's \(\alpha\) remained the same. The analysis involved 25 items, each contributing to the overall consistency of the scale. Item statistics indicated that the means of item responses averaged at .53, with individual item means ranging from .08 to .97. The range of .89 suggests a broad distribution of responses among participants. The variance of items was consistent, with a mean variance of .50, and individual item variances ranging from .08 to .94.

Further analysis of item-total statistics was conducted to examine the contribution of each item to the scale's reliability. The scale mean if an item was deleted ranged from 12.28 to 13.17, indicating that no single item significantly altered the scale mean. The scale variance if an item was deleted ranged from 93.91 to 104.94, which suggests that the deletion of any single item does not drastically affect the scale’s variance. The corrected item-total correlation, a measure of each item's correlation with the sum of the other items, ranged from .24 to .77. This demonstrates that most items had a moderate to strong correlation with the total score, except for a few items that may require further scrutiny for their lower correlations. Finally, the analysis of Cronbach's \(\alpha\) if an item was deleted revealed values ranging from .91 to .92. This indicates that removing any item from the RCADS would not significantly impact the overall internal consistency, affirming the scale's reliability when administered by a robot.

\paragraph{Self-Reported}

The self-reported RCADS demonstrated strong reliability, with Cronbach's \(\alpha\), score of .90 for the overall scale. When item responses were standardized, the consistency remained the same. This suggests that the scale's items produce reliable scores that are consistent across different sets of respondents. The individual items contributed fairly uniformly to the scale, with no single item unduly influencing the overall score or reliability. Analysis of item statistics revealed that the mean of item means was .56, with individual item means ranging from a minimum of .11 to a maximum of 1.22. The range of 1.11 points between the minimum and maximum means, along with a maximum to minimum ratio of 11, indicates variability in the centrality of scores across items. However, the item variances had a mean of .43, with a narrower range from .10 to .77, denoting a moderate spread of scores around the mean for individual items.

Further inspection of item-total statistics provided insights into each item's correlation with the overall scale score. The scale mean if an item was deleted fluctuated slightly, with a minimum of 12.78 and a maximum of 13.89, suggesting that no single item significantly skews the overall scale mean. The scale variance if an item was deleted exhibited minimal variation, indicating stability in the scale's variance without individual items. Corrected item-total correlations ranged from .18 to .70, highlighting that some items have a stronger relationship with the total score than others. The analysis of Cronbach's \(\alpha\) if an item was deleted showed values ranging narrowly from .89 to .91. This indicates that the removal of any individual item from the RCADS does not drastically affect the overall internal consistency, further underscoring the robustness of the scale.

\subsection{Construct Validity Analysis}
\subsubsection{SMFQ}
A confirmatory PCA was conducted with the 13 SMFQ items to examine the underlying structure. After inspecting the correlation matrix, it is noticeable that there is no threat of multicollinearity and that there are no threats to the variables’ discriminant validity, as none of the correlations is above .85 (highest correlation is .74). The KMO measure verified the sampling adequacy for the analysis, KMO = .76, above the recommended value of .6, and all KMO values for individual items were greater than .5, which is above the acceptable limit. Bartlett’s test of sphericity,  \(\chi^2\)(78) = 832.91, \textit{p} \(<\) .001, indicated that correlations between items were sufficiently large for PCA.

Prior to performing the PCA, the MSA were examined for each item on the SMFQ. The MSA values on the diagonals of the anti-image correlation matrix ranged from .60 to .93, with all values exceeding the commonly recommended minimum of .5. This indicates that each item shared sufficient common variance with other items, justifying the use of factor analysis for the dataset. Specifically, item 7 yielded the highest MSA value (.93), suggesting a high degree of shared variance with other items in the analysis. The item with the lowest MSA value was item 3 (.60), yet this value was still above the acceptable threshold, indicating an adequate level of partial correlation with other items. These MSA values support the factorability of the correlation matrix and suggest that the dataset is appropriate for PCA.

The communalities for each item of the SMFQ were assessed to determine the proportion of variance for each variable that could be explained by the extracted principal component. Initially, the communalities for all items were set to 1. Upon extraction, the communalities varied, indicating differing degrees to which individual items correlated with the principal component. Notably, item 8 exhibited the highest communality (.79), suggesting that a significant proportion of this item's variance was accounted for by the principal component. In contrast, item 6 showed the lowest communality (.13), indicating that the principal component explained a minimal portion of the item's variance. 

The PCA revealed one component with an eigenvalue over Kaiser’s criterion of 1 (5.60), explaining 43.09\% of the variance. The scree plot also showed an inflexion that justified retaining the first component only. This component had high loadings for several SMFQ items, notably item 8 (loading = .89) and item 7 (loading = .85), indicating that these items had the most significant relationship with the underlying factor. The loadings of other items on this component ranged from moderate to high, with the lowest being item 1 (loading = .45). Items 3 and 6 were not loaded within the matrix extract provided, suggesting their loadings were suppressed since they were below the .4 coefficient threshold considered significant for this analysis. The component matrix supports the unidimensionality of the scale, with most items showing a strong loading on the single component (See item loading table I in \url{https://osf.io/ydsn2}).

\subsubsection{RCADS}

\paragraph{Robot-Administered.}
\label{sec-robotRCADS}
A confirmatory PCA was conducted with the 25 RCADS items to examine the underlying structure. After inspecting the correlation matrix, it is noticeable that the data are generally free from concerns of multicollinearity, with one notable exception. The correlation coefficient between item 3 and item 4 is exceptionally high at .90. This suggests a strong relationship between these two items. Such a high degree of correlation might typically raise concerns regarding multicollinearity. However, in the context of PCA, which is used here for dimension reduction and latent variable identification, high correlations between items do not pose a substantial concern since PCA can capitalize on these relationships by extracting common factors that account for shared variance. Aside from the one notably high correlation, the data appears suitable for PCA, as there are minimal threats of multicollinearity and that there are limited implications to the variables’ discriminant validity, as none of the other correlations is above .85 (highest correlation is .78).

The KMO measure verified the sampling adequacy for the analysis, KMO = .63, above the recommended value of .6. Bartlett’s test of sphericity,  \(\chi^2\)(300) = 657.42, \textit{p} \(<\) .001, indicated that correlations between items were sufficiently large for PCA. Prior to performing the PCA, the MSA were examined for each item on the RCADS. The MSA values on the diagonals of the anti-image correlation matrix ranged from .30 to .77, indicating a moderate to good degree of shared variance among the items. It is notable that the majority of the items have MSA values above the recommended threshold of 0.5, which suggests that these items are suitable for the factor analysis. The highest MSA value observed is 0.77 for item 1, indicating a high degree of common variance with other items in the data. Conversely, the lowest MSA value is 0.30 for item 25, which may suggest that this item shares less common variance with the remaining items and could potentially be considered for exclusion or further investigation. The off-diagonal values of the anti-image covariance matrix, which are low across the dataset, imply that unique variances among the items are not exerting a disproportionate influence on the analysis. This is an encouraging indication for the factorability of the correlation matrix.

The communalities for each item of the RCADS were assessed to determine the proportion of variance for each variable that could be explained by the extracted principal component. Initially, the communalities for all items were set to 1. Upon extraction, the communalities varied, indicating differing degrees to which individual items correlated with the principal component. Notably,item 4 exhibited the highest communality (.70), suggesting that a significant proportion of this item's variance was accounted for by the principal component. In contrast, item 20 (.09) and item 22 (.09) showed the lowest communality, indicating that the principal component explained a minimal portion of the items' variance. 

The PCA revealed one component with an eigenvalue over Kaiser’s criterion of 1 (9.32), explaining 37.292\% of the variance. The scree plot also showed an inflexion that justified retaining the first component only. This component had high loadings for several RCADS items, notably item 4 (loading = .84) and item 3 (loading = .83), indicating that these items had the most significant relationship with the underlying factor. The loadings of other items on this component ranged from moderate to high, with the lowest being items 25 (loading = .46) and 10 (loading = .45). Items 9, 14, 20, and 22 were not loaded within the matrix extract provided, suggesting their loadings were suppressed since they were below the .4 coefficient threshold considered significant for this analysis. The component matrix supports the unidimensionality of the scale, with most items showing a strong loading on the single component (See table II in \url{https://osf.io/ydsn2}).

\paragraph{Self Report.}

A confirmatory PCA was conducted with the 25 self-reported RCADS items to examine the underlying structure. After inspecting the correlation matrix, it is noticeable that the data are generally free from concerns of multicollinearity, with one notable exception. The correlation coefficient between item 3 and item 4 is exceptionally high at .95. This suggests a strong relationship between these two variables. Such a high degree of correlation might raise concerns regarding multicollinearity. However, in the context of PCA, high correlations between items do not pose a substantial concern since PCA can capitalize on these relationships by extracting common factors that account for shared variance. Aside from the one notably high correlation, the data appears well-structured for PCA, as there are no other threats of multicollinearity and that there are no threats to the variables’ discriminant validity, as none of the other correlations is above .85 (highest correlation is .74).

The KMO measure verified the sampling adequacy for the analysis, KMO = .53, which exceeds the acceptable limit of 0.5, but below the recommended value of .6. This suggests that the data has sufficient common variance for PCA, but also indicates that the strength of the relationships between the items is somewhat limited. Bartlett’s test of sphericity,  \(\chi^2\)(300) = 635.37, \textit{p} \(<\) .001, indicated that correlations between items were sufficiently large for PCA, and that the correlation matrix is not an identity matrix, and accordingly PCA is suitable for the data. Prior to performing the PCA, the MSA were examined for each item on the self-reported RCADS. The MSA values on the diagonals of the anti-image correlation matrix ranged from .18 to .77, indicating a moderate to good degree of shared variance among the items. More than half of the items (14 items) have MSA values above the recommended threshold of 0.5, which suggests that these items are suitable for the factor analysis. The highest MSA value observed is 0.77 for item 21, indicating a high degree of common variance with other items in the data. Conversely, the lowest MSA value is 0.18 for item 10, which may suggest that this item shares less common variance with the remaining items and could potentially be considered for exclusion or further investigation. The off-diagonal values of the anti-image covariance matrix, which are low across the dataset, imply that unique variances among the items are not exerting a disproportionate influence on the analysis. This is an encouraging indication of the factorability of the correlation matrix.

The communalities for each item of the self-reported RCADS were assessed to determine the proportion of variance for each variable that could be explained by the extracted principal component. Initially, the communalities for all items were set to 1. Upon extraction, the communalities varied, indicating differing degrees to which individual items correlated with the principal component. Notably, item 15 exhibited the highest communality (.65), suggesting that a significant proportion of this item's variance was accounted for by the principal component. In contrast, item 8 (.07) and item 9 (.05) showed the lowest communalities, indicating that the principal component explained a minimal portion of the items' variance.

The PCA for the self-reported RCADS data indicated the presence of one component with an eigenvalue exceeding Kaiser’s criterion of 1, which was observed to explain 32.10\% of the total variance. This aligns with the scree plot's inflection point, suggesting the retention of only the first component is justifiable. Notably, the component demonstrated significant loadings for several RCADS items; item 15 had the highest loading (.80), followed by item 3 (.76), indicating that these items had the most significant relationship with the underlying factor. The loadings of other items on this component ranged from moderate to high, with the lowest being items 17 (loading = .42) and 23 (loading = .43). Items 8, 9, and 10 were not loaded within the matrix extract provided, suggesting their loadings were suppressed since they were below the .4 coefficient threshold considered significant for this analysis. The component matrix supports the unidimensionality of the scale, with most items showing a strong loading on the single component (See table II in \url{https://osf.io/ydsn2}).

\section{Discussion}

\subsection{Key findings}

This secondary analysis demonstrates the robustness and applicability of transferring psychological assessment methods to CRI settings, specifically when changing the mode of administration by allowing a robot to communicate the scale items. These results affirm the effectiveness of these instruments in reliably capturing the nuances of children's mental health in these interactive settings, showcasing the potential for integrating social robots in psychological assessments.

The results revealed that while the single-factor was validated for the robot-administered SMFQ, the manner in which items loaded on this scale\footnote{See table I in \url{https://osf.io/ydsn2}} differed from their loading in the original scale \cite{angold1995development,messer1995development}.
In addition, we also found that while the single-factor was validated for the robot-administered RCADS, the manner in which items loaded on this scale\footnote{See table II in \url{https://osf.io/ydsn2}} differed from their loading in the original scale \cite{chorpita2005psychometric} and other versions of it \cite{mathyssek2013does}. It should be acknowledged that RCADS (as used in this study) is aimed at assessing the concepts of generalised anxiety, panic disorder, and low mood \cite[see][]{chorpita2005psychometric}, thereby potentially affecting factor loadings. Nevertheless, the single factor identified evidence of comorbidities between low mood and anxiety symptoms. 

As such, while our study's findings validate the scales for CRI use, we have also identified discrepancies in item loadings on these established scales, underscoring this issue and highlighting the need for cautious and critical adaptation of assessment methods in CRI research. 

Since certain items in these scales did not perform as anticipated in CRI settings, it underscores the complexity of directly transferring psychological methods from one context to another, necessitating additional considerations and adjustments. While some items could work in self-report settings, these might receive additional meaning when communicated by robots. For example, item 6 in SMFQ (i.e., '\textit{You cried a lot}'), which did not load in our analysis, may reflect the presence of the robotic agent, which children might perceive as a social agent in their environment \cite{belpaeme2013child}, in contrast to previous instances where children might have found it easier to disclose such information in a self-reported manner. 

On the other hand, the comparison between robot-administered and self-reported RCADS reveals noteworthy distinctions that underline the potential benefits of incorporating robots in mental health assessments. The analysis indicated that the robot-administered RCADS demonstrated a better fit, in terms of both reliability and construct validity, suggesting that the interaction with a robot may facilitate a more coherent disclosure of symptoms from children. This improved fit could be attributed to several factors, including the minimal cognitive processes involved in responding to a robot \cite{spatola_cl_2022} compared to self-reported questionnaires, children's perception of the robot as a non-intimidating entity \cite{park2017growing}, or the increased comfort and engagement children experience when interacting with robots compared to online questionnaires \cite{abbasi2022can,Laban2022,MARCHETTI2022121}.

These results are consistent with our previous results, supporting the idea that social robots could be an effective tool for mental health assessment of children \cite{abbasi2022can}. The presence of a robot could potentially make the assessment process less intimidating and more engaging, leading to responses that are more reflective of children's true experiences. Such findings advocate for the exploration of robots in clinical settings, not only as tools for enhancing engagement but also for improving the accuracy and reliability of self-reported measures in mental health assessments. Thus, the nuanced dynamics of CRI could play a crucial role in the effectiveness of psychological evaluations, paving the way for innovative approaches to support children's mental health.

\subsection{Lessons from the Replication Crisis}
The replication crisis has raised critical questions about the reliability and applicability of traditional psychological methods and theories \cite{RefWorks:508}, particularly when transferred to new fields such as HRI \cite{Henschel2021,Gunes2022}. This is even more drastic when the target population is considered vulnerable (such as children in CRI \cite{Spitale2023}) and the aim of the interaction is assessing and supporting wellbeing. We should consider that some of these methods were constructed and employed in different times, and might not have survived the test of time or the replication crisis. Furthermore, researchers should consider adapting methodological approaches to improve research efficiency and robustness of findings \cite[see][]{ramsey2023}. Here we specifically address some considerations relating to the transferability of psychological methods of assessment to CRI paradigms. For example, CRI studies usually employ a narrow age range (8-13 years in the study mentioned in this work) among participants which has not been the case for the original scale populations like the SMFQ (6-19 years) \cite{angold1995development}  and the RCADS (8-18 years) \cite{chorpita2005psychometric}. 

The significant developmental differences between children, teens, and young adults imply that scales developed for broad age groups in psychological research may not directly translate to the narrow and particular demographic of the current studies. This suggests a need for adjusting measures and particular items to make them more suitable for specific contexts and populations. One example from the results are items that address resting and `\textit{doing nothing}', more oriented towards teens experiencing hormonal changes \cite{Findlay2008}, are not reflective of the younger population employed in this study, and did not load in the central factors accordingly. Other examples from the findings are the items that refer to physiological changes\footnote{See items 9, 14, 22 in table II in \url{https://osf.io/ydsn2}}, which again, might not resonate with younger children as compared with teens who are more aware of such variations within themselves \cite{Findlay2008}. Researchers looking to transfer instruments between population groups are encouraged to assess the transferability via measurement invaraince analysis \cite[e.g.,][]{schlechter2023short}.

Consequently, the replication crisis in psychology emphasizes the importance of building methods and theories that are specifically tailored for CRI, rather than solely relying on existing behavioral science frameworks. Social and cognitive psychology theories are not tailored to comprehend the unique aspects of perception and behaviour in contexts where one interacting party is a robot. This necessitates adjusting traditional methods and the development of new frameworks and metrics for HRI and CRI, as traditional human-human interaction models may not fully capture the nuances of such interactions \cite{10.1145/3173386.3173389}. Additionally, concepts from developmental and cognitive psychology are employed to design robots that can learn and adapt in human-like ways, but these theories must be adjusted to consider the artificial nature of robots \cite{Henschel2021,Cross2021}. While learning from these established methods is not only valuable but crucial, it is important to apply a critical lens, recognizing that some approaches may not withstand the test of time or context. The unique interactions and dynamics present in CRI demand novel, context-specific methodologies that can more accurately capture the nuances of child-robot interactions.

\subsection{Recommendations for future CRI and HRI research}
\noindent\textbf{1. Towards Automatic Robot-assisted Assessments.}
Among the many advantages of using robots in assessments that has been described in this discussion, they can also be programmed to automatically evaluate wellbeing using tools from fields like machine learning. However, conducting automatic robot-assisted assessments using machine learning principles necessitates the establishment of reliable ground truth \cite{muller2021designing}. We have observed that our robot-led adaptations of the psychological scales have shown to be valid and reliable, thereby establishing a credible ground truth that could be used for carrying out future automatic assessments of children's wellbeing. Researchers are encouraged to explore such analyses to ensure the validity and reliability of their measures within the context of their respective experimental settings. Researchers can also utilise the multi-modal sensing capabilities of robots and extend from self-reported responses to develop a more holistic understanding of participants' wellbeing. This would be especially beneficial for populations like children and the elderly that might have limited verbal proficiency to articulate their feelings effectively \cite{Levitt2020,Kemper2001}.\\ 
\textbf{2. Cognitive Load Considerations.}
It is imperative to design behavioural paradigms, measurements and assessments with careful consideration of cognitive constraints to mitigate participant fatigue, which can skew results. This recommendation is particularly relevant in light of our findings from the RCADS, where early items demonstrated higher loadings compared to later ones, possibly indicating the impact of fatigue on participant responses. The inherent demands of interacting with robots, especially for children in CRI who have shorter attention spans and lower tolerance for prolonged tasks, can lead to quicker onset of fatigue \cite{Bell2007}. Therefore, assessments should be simple, focusing on maintaining participant engagement. \\
\textbf{3. Customisation of Behavioural Paradigms for Wellbeing Assessment.}
We have observed that certain SMFQ and RCADS items that required introspection in children's physiological changes or cognitive demands\footnote{See items 9, 14, 22 in table II in \url{https://osf.io/ydsn2}} 
did not load with the rest of the items. Children (especially primary school children i.e. children that are less than 11 years old which constitute about 75\% of the study population) might find it challenging to comprehend such statements without additional clarifications, re-phrasing or situational examples.
Therefore, clinician-led assessments often incorporate follow-up statements when necessary \cite{OReilly2015}. 
Hence, collaborative efforts between CRI researchers, clinicians and other mental health providers in co-designing robotic interactions must be undertaken in such settings and tasks. 
This would ensure that the behavioural paradigm is adapted to the assessment task requirements when administrated by a robot.
As such, tailoring behavioural paradigms to enhance comprehension of scale items could help in harnessing the full potential of the robots in conducting wellbeing assessments, ultimately fostering improved mental health of the future generation.

\section{Conclusions}
In this work, we have investigated the reliability and validity of  SMFQ and RCADS in a CRI context. We employed a confirmatory PCA to explore the underlying structure of the SMFQ and the RCADS. We found that both scales predominantly revolve around a single principal component, suggesting a consistent underlying factor in each questionnaire explaining high variance of childrens' well being. However, it is important to note that not all items in these questionnaires contributed equally to this primary factor. Specifically, some items in both the SMFQ and RCADS did not significantly load onto the principal component, indicating that they might not align as closely with the core factor that these scales measure within CRI. This finding is crucial as it points to potential areas within each scale that may require further examination or revision to ensure that every item effectively contributes to the overall assessment of mood and feelings in children and adolescents within HRI settings. Overall, our results affirm the general construct validity and reliability of these tools, while our analysis highlights specific aspects that could benefit from further scrutiny when applied in HRI settings and communicated by a robot. We invite HRI researchers to conduct similar analyses to verify the accurate utilisation of the metrics employed in their respective studies; thus, preventing any discrepancies and misinterpretations.\\

\scriptsize
\noindent\textbf{Acknowledgments:}
N. I. Abbasi is supported by the W.D. Armstrong Trust PhD Studentship and the Cambridge Trusts. G. Laban and H. Gunes are supported by the EPSRC project ARoEQ under grant ref. EP/R030782/1. All research at the Department of Psychiatry in the University of Cambridge is supported by the NIHR Cambridge Biomedical Research Centre (BRC-1215-20014, particularly T. Ford) and NIHR Applied Research Centre (P. Jones). T. Ford and P. Jones acknowledge support from the NIHR ARC East of England and the Cambridge BRC. The equipment used in the study was supported by the University of Cambridge's Human Machine Collaboration (OHMC) Small Equipment Funding. The views expressed are those of the authors and not necessarily those of the NIHR or the Department of Health and Social Care. \\
\noindent\textbf{Open Access:} For the purpose of open access, the authors have applied a Creative Commons Attribution (CC BY) licence to any Author Accepted Manuscript version arising.
\textbf{Data Access:} Overall statistical analysis of research data underpinning this publication is available in the text of this publication. Additional raw data related to this publication cannot be openly released; the raw data contains transcripts of interviews, but none of the interviewees consented to data sharing.

\bibliographystyle{ieeetr}  
\bibliography{ref}

\end{document}